\documentclass[twocolumn,prb,showpacs,superscriptaddress]{revtex4}

\usepackage{amsmath}
\usepackage{graphicx}
\usepackage{amssymb}
\usepackage{color}
\usepackage{epsfig}
\usepackage{epstopdf}
\usepackage{dcolumn}
\usepackage{bm}

\begin{document}

\title{Hybrid Spin Noise Spectroscopy and the Spin Hall Effect}

\author{Valeriy A. Slipko}
\affiliation{Department of Physics and Astronomy, University of South Carolina, Columbia, South Carolina 29208, USA}
\affiliation{ Department of Physics and Technology, V. N. Karazin
Kharkov National University, Kharkov 61077, Ukraine}

\author{Nikolai A. Sinitsyn}
\affiliation{Theoretical Division, Los Alamos National Laboratory, Los Alamos, New Mexico 87545, USA}

\author{Yuriy V. Pershin}
\email{pershin@physics.sc.edu}
\affiliation{Department of Physics and Astronomy, University of South Carolina, Columbia, South Carolina 29208, USA}

\begin{abstract}
Here we suggest a novel hybrid spin noise spectroscopy technique, which is sensitive to the spin Hall effect. It is shown that, while the standard spin-spin correlation function is not sensitive to the spin Hall effect, spin-transverse voltage and transverse voltage-voltage correlation functions provide the missing sensitivity being linear and quadratic in
the spin Hall coefficient, respectively. The correlation between transverse voltage and spin fluctuations appears as a result of spin-charge coupling fundamental for the spin Hall effect. We anticipate that the proposed method could find applications in the studies of spin-charge coupling in semiconductors.
\end{abstract}

\pacs{72.25.Dc,72.20.Jv,71.70.Ej}

\maketitle

\textit{Introduction.}---The spin noise spectroscopy (SNS) is an emergent experimental technique that has been used or suggested to measure spin-related parameters of different materials and systems \cite{Oestreich2005a,Muller08a,Crooker09a,Muller10a,glazov2011theory,Li12,Glazov13a}.
Experimentally, the SNS can be realized using a linearly polarized laser beam that is sent through the sample \cite{Muller10a}. Due
to the Faraday effect, the laser beam polarization plane rotates depending on the electron spins inside the illuminated region of the sample.
At equilibrium, in spin-unpolarized systems, the time-averaged Faraday rotation angle is zero. The mesoscopic fluctuations of the Faraday rotation angle are used to find spin-spin correlation function. Typically, its Fourier transform, called the noise power spectrum, contains a peak. The peak's parameters (such as position and width) provide important information about spin-related system parameters, such as the g-factor and spin relaxation time. Recently, the SNS has been extended to probe equilibrium \cite{pershin13a} and nonequilibrium  \cite{Li13a} {\it transport} characteristics of conducting electrons.

Being a minimally invasive technique, the SNS provides information  without virtually any dissipation of energy in the system.
The SNS is particularly powerful at systems that are difficult to study by pump-probe techniques. For example,  it achieved the unprecedented resolution of the relaxation spectrum of the central spin in self-assembled InGaAs quantum dots\cite{Li12}, which exposed the effect of the quadrupole coupling of nuclear spins on the decoherence and relaxation of the central spin qubit\cite{Sinitsyn12}.

 In this Letter, we propose an experimental setup that probes spin fluctuations for the goal of exploring the physics of the spin Hall effect \cite{Dyakonov71a,hirsch1999spin}.
In this effect, the spin current is generated in the direction perpendicular
to the electric field applied to a non-magnetic electron system \cite{Zhang00a,rashba-prl,hankiewicz2009spin}. While the spin Hall effect has been observed in numerous experiments, its quantitative characterization is usually not a simple task because spin is generally not a conserved degree of freedom. In addition, converting spin currents into the voltage signal needs application of ferromagnetic contacts or inhomogeneous doping \cite{Nikolic09a,pershin07c}, which is not straightforward for many semiconductor structures. Here we show that the SNS can provide an alternative approach to measure parameters of the spin Hall effect, avoiding many difficulties of the previously used experimental techniques.

\begin{figure}[b]
\centering
\includegraphics[width=0.9\linewidth]{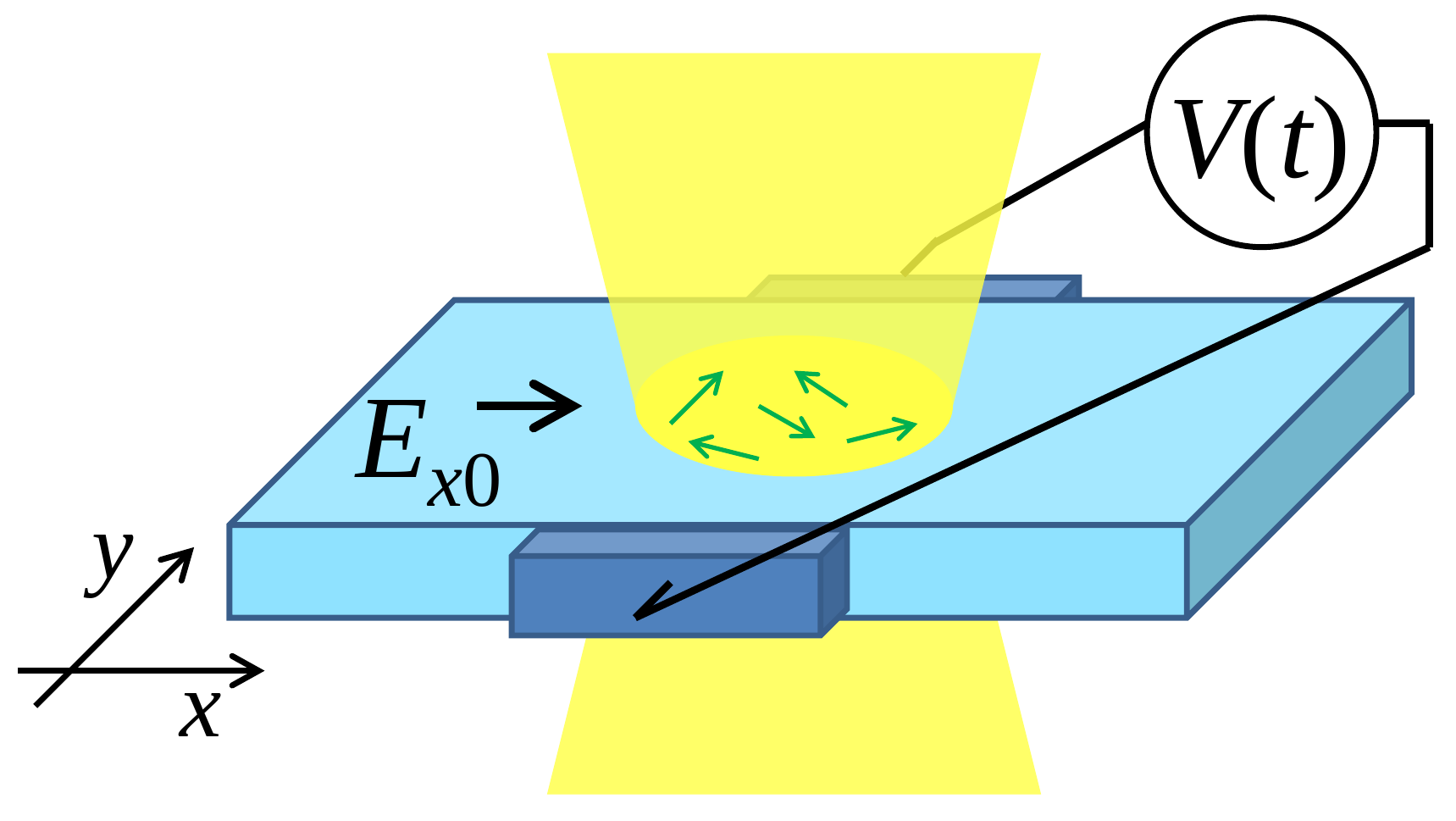}
\centering \caption{Hybrid spin noise spectroscopy of the spin Hall Effect. Correlations between spin (measured optically) and transverse voltage (measured electronically) fluctuations provide information regarding the spin Hall effect.}
\label{fig0}
\end{figure}

In a typical semiconductor material, such as GaAs, the spin Hall effect is relatively weak.
It cannot appreciably modify spin fluctuations observed by the standard optical SNS setup.
In order to overcome this difficulty, we suggest a method of {\it hybrid spin noise spectroscopy}, which is based on a simultaneous analysis of spin and transverse voltage fluctuations, as shown in Fig. \ref{fig0}. In this setup, one can experimentally determine
spin-transverse voltage (SV) and transverse voltage-transverse voltage (VV) correlation functions which are sensitive to the spin Hall coefficient.
Opposite to the conventional Hall effect, the spin Hall effect in homogeneous systems is not accompanied by any transverse voltage on average \cite{pershin07c}. However, as we will demonstrate below, in the spin Hall regime the spin fluctuations are dressed by charge dipoles that are responsible for the transverse voltage fluctuations. Therefore, the transverse voltage fluctuations correlate with the spin fluctuations and their strength is proportional to the spin Hall coefficient.

 According to the theory presented below, SV and VV correlation functions can be written as
\begin{eqnarray}
\left< S_z(0)V(t)\right> \sim \gamma E_{x0} e^{-\nu_st}, \label{eq_res1} \\
\left< V(0)V(t)\right> \sim \left(\gamma E_{x0}\right)^2 e^{-\nu_st}, \label{eq_res2}
\end{eqnarray}
where $\gamma$ is the dimensionless spin Hall effect coefficient describing deflection of spin-up (+) and spin-down (-) electrons, $E_{x0}$ is the longitudinal (applied) electric field and $\nu_s$ is the spin relaxation rate. This is the main result of our theory. In the following, we shall derive Eqs. (\ref{eq_res1}) and (\ref{eq_res2}) using the standard two-component drift-diffusion equation approach \cite{Yu02a,Tse05a,pershin07b}, which is appropriate for the description of the extrinsic spin Hall effect in 2D electron systems. First of all, we develop a theory of coupled spin-charge fluctuations and then use its results to find SV and VV correlation functions.

\textit{Dynamics of spin fluctuations.}---Within the two-component drift-diffusion equation approach \cite{Yu02a,Tse05a,pershin07b}, the charge and spin degrees of freedom are described by the following equations:
\begin{eqnarray}
e\frac{\partial n^{\pm}}{\partial t}
&=& \frac{\partial j^{\pm}_i}{\partial x_i}
-\frac{e}{2\tau_{s}}\left(n^{\pm}-n^{\mp}\right),
\label{cont2D} \\
 j^{\pm}_i &=& \lambda_{ij}^{\pm}E_{j}+eD\frac{\partial n^{\pm}}{\partial x_{i}},
~i=1,2,
\label{current2D}
\\
\lambda^{\pm} &=& en^{\pm}\mu\begin{pmatrix}1 & \mp\gamma \\
\pm\gamma & 1 \\
\end{pmatrix},
 \label{lambda_matrix}
\\
\frac{\partial E_i}{\partial x_{i}} &=& \frac{e}{\varepsilon\varepsilon_0}\left( N_0-n\right),
\label{pua2D}
\end{eqnarray}
where $-e$ is the electron charge, $n^{\pm}$ is
the volume density of spin-up (sign "+") and spin-down electrons (sign "-"), $j_{i}^{\pm}$ is $i$th component of the current density of spin-up (spin-down) electrons, $\tau_{s}=\nu_s^{-1}$ is the spin
relaxation time, $\lambda_{ij}^{\pm}$ is the tensor of spin-up (spin-down) conductivity, $\mu>0$
is the mobility, $D$ is the diffusion coefficient, $\varepsilon$ is
the permittivity of the bulk, $N_0$ is the background volume charge density.
We assume the summation over repeated twice indexes, which run from 1 (x-axis) to 2 (y-axis). The spin Hall effect is described by the non-diagonal  terms in the conductivity matrix Eq. (\ref{lambda_matrix}). Eq. (\ref{cont2D}) is the continuity relation that takes into
account spin relaxation and Eq. (\ref{pua2D}) is the Poisson
equation. Eq. (\ref{current2D}) is the expression for the
current density   which includes drift, diffusion
and spin Hall effect components.


For the sake of simplicity, we assume homogeneous $x$-component
of the electric field in both $x$ and $y$ directions, and homogeneous charge and current densities in the $x$ direction. The latter assumption is compatible with Fig. \ref{fig0} setup, which does not provide spacial resolution. Combining Eqs. (\ref{cont2D}) and (\ref{current2D}) for
different spin components we can get the following equations for the
electron density $n=n^{+}+n^{-}$ and the spin
density imbalance $P=n^{+}-n^{-}$:
\begin{eqnarray}
\frac{\partial n}{\partial t}&=&\frac{\partial}{\partial y} \left[ \mu  E_y n +
D\frac{\partial n}{\partial y} +\gamma  \mu E_x P
\right] \label{neq},\\
\frac{\partial P}{\partial t}&=&\frac{\partial}{\partial y}
\left[ \mu  E_y P + D \frac{\partial P}{\partial y} +\gamma  \mu
E_x n\right]-\nu_s P. \label{Peq}
\end{eqnarray}
In the absence of fluctuations, the distribution of spin polarization (the steady-state solution of Eqs. (\ref{pua2D})-(\ref{Peq})) is well known. It consists in spin accumulation on the sample boundaries (in the opposite directions on the opposing boundaries)
and zero spin polarization in the bulk \cite{Kato04a}. Moreover, in the bulk, $E_y=0$ and $n=N_0$.

Let us now consider a small fluctuation. We linearize Eqs. (\ref{pua2D})-(\ref{Peq}) about
the steady-state solution $n=N_0$, $P=0$ and $E_y=0$ and obtain the following linearized equations
\begin{eqnarray}
\frac{\partial \tilde n}{\partial t}&=&D\frac{\partial^2 \tilde n}{\partial y^2} -\nu\tilde n+
\gamma\mu E_{x0}\frac{\partial P}{\partial y}
 \label{neqlin}, \\
\frac{\partial P}{\partial t}&=&D\frac{\partial^2 P}{\partial y^2} -\nu_s P+
\gamma\mu E_{x0}\frac{\partial \tilde n}{\partial y}
, \label{Peqlin}\\
\frac{\partial E_y}{\partial y
}&=&-\frac{e}{\varepsilon\varepsilon_0}\tilde n,
\label{puaeqlin}
\end{eqnarray}
where $\nu=\mu N_0 e/(\epsilon\epsilon_0)$ and $n=N_0+\tilde n$.
The general solution of Eqs. (\ref{neqlin})-(\ref{Peqlin}) can be written as
\begin{widetext}
\begin{eqnarray}
P&=&e^{-(\nu+\nu_s)t/2}\int_{-\infty}^{+\infty}dk e^{iky-Dtk^2}
\left(P_{0}(k)\cosh(Qt)+\left[\frac{1}{2}(\nu-\nu_s)P_{0}(k)+ik\gamma\mu
E_{x0}\tilde n_{0}(k)\right]\frac{\sinh(Qt)}{Q}\right),~~~~
\label{Pgs}
\\
\tilde n&=&e^{-(\nu+\nu_s)t/2}\int_{-\infty}^{+\infty}dke^{iky-Dtk^2}
\left(\tilde n_{0}(k)\cosh(Qt)+\left[ik\gamma\mu
E_{x0}P_{0}(k)-\frac{1}{2}(\nu-\nu_s)\tilde n_{0}(k)\right]\frac{\sinh(Qt)}{Q}\right),
\label{ngs}
\end{eqnarray}
\end{widetext}
where $\tilde n_{0}(k)=\int_{-\infty}^{+\infty}dy\exp(-iky)\tilde n(y,0)/(2\pi)$ and $P_{0}(k)=\int_{-\infty}^{+\infty}dy\exp(-iky) P(y,0)/(2\pi)$ are functions determined by the initial distributions of charge density and imbalance, and
$Q(k)=\sqrt{(\nu-\nu_s)^2/4-(\gamma\mu E_{x0}k)^2}$.

In what follows we shall carefully study the solution of Eqs. (\ref{neqlin})-(\ref{puaeqlin}) using the
approximation of Dirac delta function as the initial spin density imbalance, namely,
\begin{equation}
\tilde n(y,t=0)=0,~~P(y,t=0)=P_0\delta(y).
\label{IC}
\end{equation}
In particular, we will focus on the limiting case when $\nu_{s}/\nu\ll 1$ and $(\gamma\mu E_{x0})^2/(\nu D)\ll 1$.
According to the definition given below Eq. (\ref{puaeqlin}), $\nu^{-1}$ is a characteristic charge relaxation time.
Typically, in semiconductors, the process of electron spin relaxation is much slower then that of the charge relaxation and thus the first inequality is
commonly valid. The validity of the second inequality can be straightforwardly demonstrated for the case of typical GaAs parameters. Taking~\cite{pershin09b} $\mu=8500$~cm$^2$/(Vs),
$D=55$~cm$^2$/s, $\varepsilon=12.4$, $\nu_s=10^{8}$~s$^{-1}$, $\gamma=10^{-3}$,
$N_0=10^{16}$~cm$^{-3}$, $E_{x0}=100$~V/cm, we find
 $\nu_{s}/\nu=0.81\cdot 10^{-5}$ and
 $(\gamma\mu E_{x0})^2/(\nu D)=1.1\cdot 10^{-9}$.

In this case the solution of Eqs. (\ref{neqlin})-(\ref{puaeqlin}) takes the following uncomplicated form
\begin{eqnarray}
\tilde n(y,t)&=&-P_0\frac{\gamma\mu E_{x0}}{4\nu\sqrt{\pi(Dt)^3}}
\left(e^{-\nu_s t}-e^{-\nu t}\right)
y e^{-\frac{y^2}{4Dt}},~~~~
\label{nss}
\\
P(y,t)&=&P_0
 e^{-\nu_s t}\frac{e^{-y^2/(4Dt)}}{\sqrt{4\pi Dt}},
\label{Pss}\\
 E_y(y,t)&=&-P_0\frac{\gamma E_{x0}}{N_0}
 \left(e^{-\nu_s t}-e^{-\nu t}\right)
 \frac{e^{-y^2/(4Dt)}}{\sqrt{4\pi Dt}}.
\label{Eyss}
\end{eqnarray}
According to Eq. (\ref{nss}), the spin fluctuation is dressed by a charge dipole. This dipole is formed on the short $\nu^{-1}$ time scale and then slowly decays coherently with the spin polarization on the time scale of $\nu_s^{-1}$. An internal electric field (Eq. (\ref{Eyss})) is induced by the charge dipole.

\begin{figure}[t]
\centering
\includegraphics[width=0.9\linewidth]{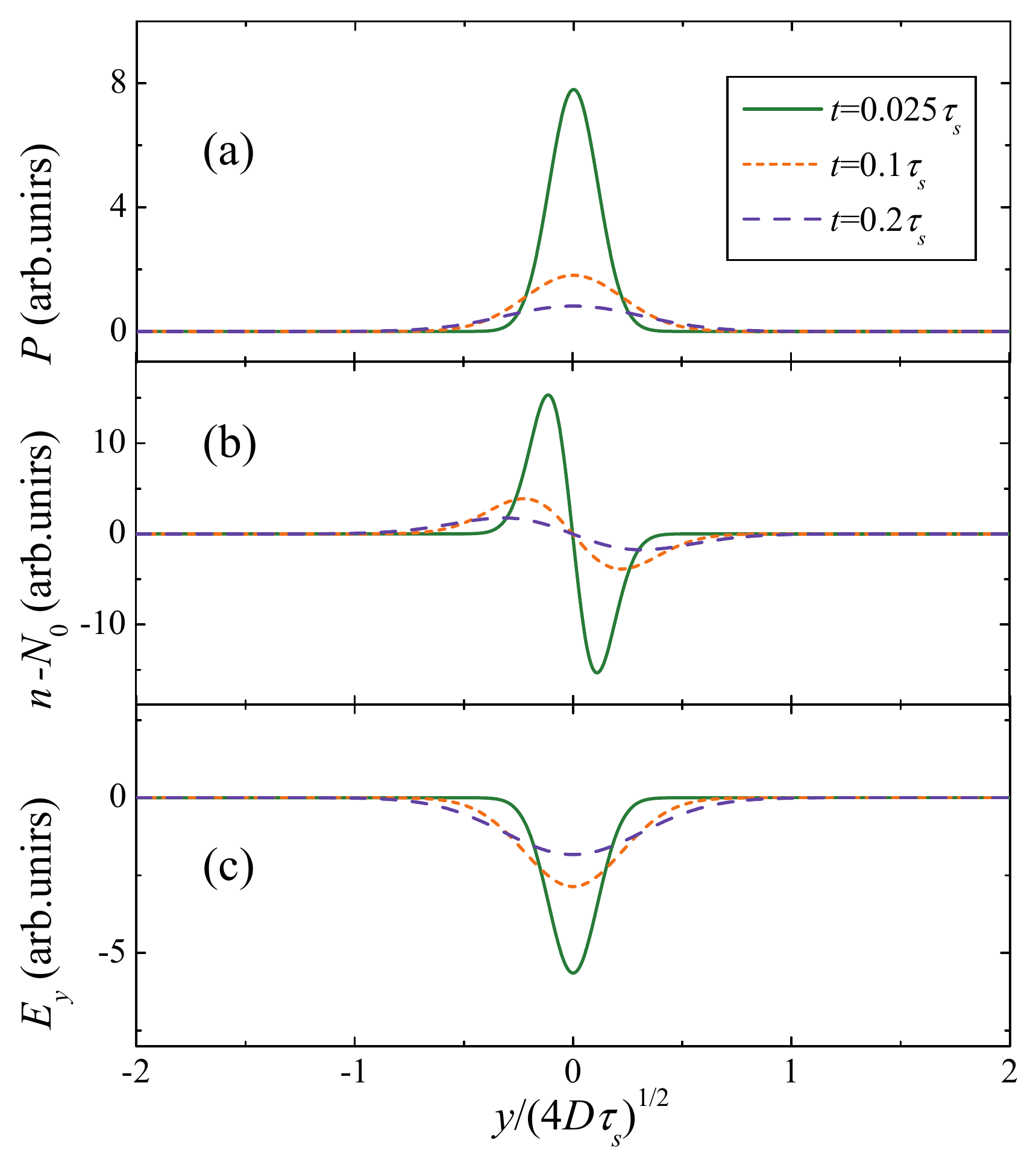}
\centering \caption{ Dynamics of spin density fluctuation: the spin density imbalance (a), excess electron density (b) and transverse electric field (c) at different moments of time indicated in (a). This plot was obtained using the parameter value $\nu / \nu_s=100$.}
\label{fig2}
\end{figure}

Moreover,  the evolution of the spin fluctuation (Eq. (\ref{Pss})) is a simple diffusion combined with exponential decay without any drift.  Fig. \ref{fig2} illustrates the solution given by Eqs. (\ref{nss})-(\ref{Eyss}) showing this nontrivial feature. One can naively expect that the spin Hall effect should lead to a lateral drift of spin fluctuations. However, the lateral drift is second order in $\gamma$. Additionally, one can notice that the built-in electric field (due to the charge dipoles) prevents the drift of spin fluctuations. This effect is also weak and plays a role only in the case of significant spin polarization. Moreover, when multiple spin fluctuations are present, spin fluctuations of different sign attract each other and prevent each other from the lateral drift on average.

The polarization of charge dipole in Fig. \ref{fig2}(b) is determined by the general drift direction of spin-up electrons (from right to left). Moreover, the emergence of the charge dipole can be used to test fluctuations through the transverse voltage. Indeed, even a local electric field (such as that given by Eq. (\ref{Eyss})) creates a global voltage drop across the system in the transverse direction. The transverse voltage  drop $V_t$  is given by
\begin{equation}
V(t)=-\int_{-\infty}^{+\infty}E_y dy=P_0
\frac{e\gamma\mu E_{x0}}{\epsilon\epsilon_0(\nu-\nu_s)}
 \left(e^{-\nu_s t}-e^{-\nu t}\right).
\label{Vss}
\end{equation}
According to Eq. (\ref{Vss}), the transverse voltage exhibits a sharp rise on the time scale of $\nu^{-1}$ followed by exponential decay determined by $\nu_s$.

\textit{Correlation functions.}---We now apply the above theory to the experimental setup shown in Fig. \ref{fig0}. First of all, we note that the Faraday rotation setup provides an averaged space distribution of the Faraday rotation angle
$\theta (\mathbf{r},t)=\kappa S_z(\mathbf{r},t)$ according to \cite{pershin13a}
\begin{equation}
\bar{\theta}(t) =\frac{1}{I_0} \int\limits_A I(\mathbf{r})\theta (\mathbf{r},t)\textnormal{d}\mathbf{r} \approx \frac{\kappa}{L_y}\int\limits_0^{L_y} S_z (y,t)\textnormal{d}y, \label{av_theta}
\end{equation}
$\kappa$ is a constant that couples $z$-component of spin polarization density with a local value of Faraday rotation angle, $I_0$ is the integrated laser beam intensity (power), $I(\mathbf{r})$ is the space distribution of the beam intensity, which is approximated by a constant in Eq. (\ref{av_theta}), and $L_y$ is the system width.

We also assume that at the initial moment of time $t=0$ (that can be arbitrarily selected) the vector of spin polarization density per unit area  is given by a vector of continuous random variables $\mathbf{S}(\mathbf{r},0)=\boldsymbol{\xi}(\mathbf{r})$ such that $\langle  \xi_i(\mathbf{r}) \rangle =0$ and $\langle \xi_i(\mathbf{r}) \xi_j(\mathbf{r'}) \rangle=\lambda \delta(\mathbf{r}-\mathbf{r'})\delta_{ij}$, where $\langle .. \rangle$ denotes averaging over different realizations, $i,j=x,y,z$, and $\lambda$ is a parameter describing the strength of spin fluctuations. Using statistical considerations~\cite{Reif65a}, one can find that in non-degenerate electron gas $\lambda=n_{2D}/4$, where $n_{2D}= N_0L_z$ is 2D electron density. In the case of degenerate electrons, $\lambda=D(\epsilon_F)L_zk_BT/4$, where $D(\epsilon_F)$ is the energy density of states at the Fermy energy $\epsilon_F$.  Furthermore, we approximate the noise correlations as $\langle S_z(y,0)S_z(y',0) \rangle=\lambda \delta(y-y')/L_x$ to make them compatible with our effective 1D spin fluctuations theory.

In the linear approximation, the transverse voltage at time $t$ (for a given initial spin polarization distribution) can be written as
\begin{equation}
V(t)= \int \frac{2S_z(y,0)}{L_z}
\frac{e\gamma\mu E_{x0}}{\epsilon\epsilon_0(\nu-\nu_s)}
\left[e^{-\nu_st}-e^{-\nu t} \right]\textnormal{d}y+\chi(t),
\label{Vtr}
\end{equation}
where $\chi(t)$ describes all the additional noise created in the system between $0$ and $t$, which does not correlate with $S_z(y,0)$. Assuming a long time limit, one can thus find the experimentally measurable correlation function of optical and voltage signals
\begin{equation}
\langle \theta(0) V(t) \rangle \approx \frac{2\kappa \lambda}{L_{x}L_z}\frac{\gamma E_{x0}}{N_0}e^{-\nu_st}, \label{SVcorr}
\end{equation}
which is proportional to SV correlation function.

Similarly, one can find VV correlation function. For this purpose, as the initial moment of time, we consider a time which is much longer than $\nu^{-1}$ but much shorter than $\nu_s^{-1}$. This procedure properly takes into account the fast processes of charge equilibration in the system.  VV correlation function found this way can be written as
\begin{equation}
\langle V(0) V(t)\rangle= \frac{4\lambda L_y}{L_xL^{2}_z} \left( \frac{\gamma E_{x0}}{N_0}\right)^2 e^{-\nu_st}. \label{VVcorr}
\end{equation}

Let us now estimate the magnitude of VV correlations. For this purpose, we compare the thermal noise per unit frequency with the peak amplitude in the Fourier transform of Eq. (\ref{VVcorr}) near its maximum in GaAs. Taking $L_x=20\mu$m, $L_y=20\mu$m, $L_z=1\mu$m, $N_0=10^{17}$cm$^{-3}$, $\tau_s=10^{-8}$s, $\mu=8500$cm$^2$/(Vs), $\gamma=10^{-3}$, $E_{x0}=100$V/cm, $T=293$K one can find $\sqrt{2Rk_BT} \approx 0.8$nV/$\sqrt{\textnormal{Hz}}$ and
$\sqrt {2V(0)^2\tau_{s}}\approx 2\cdot 10^{-12}$V/$\sqrt{\textnormal{Hz}}$. Therefore, in materials with weak SO coupling such as GaAs, the main focus should be placed on studies of SV correlations, while materials with strong SO coupling are better positioned for measurements of VV correlations.

In order to distinguish the physical correlator from the other extrinsic background noise sources, one can apply the approach of Ref. \onlinecite{Crooker09a}. In the frequency domain, correlators such as (\ref{eq_res1}) appear as Lorentzian peaks in the noise power spectrum. The background noise, such as due to the shot noise of photons or  thermal voltage fluctuations, can be orders of magnitude stronger than the useful signal. However, the position of the peak due to the spin dynamics is sensitive to the applied external in-plane magnetic field, which shifts its maximum to the Larmor frequency. Hence, by measuring the noise power in a weak and then in strong magnetic fields, and then subtracting results, one effectively cancels the background contributions and obtains a physical correlator.

\textit{Discussion.}---
We have suggested the method of hybrid spin noise spectroscopy, which can be used to directly probe spin-charge coupling as in the case of the spin Hall effect. We have demonstrated that the spin Hall effect is manifested in the transverse voltage fluctuations, which could be probed by our setup. In particular, we found that the strength of the spin-voltage correlator is proportional to the spin Hall angle, and hence its measurements can be used to estimate this coefficient, as well as other parameters, such as the spin relaxation time. This is a  potentially interesting approach to detect and study the spin Hall effect as it is well known that on average the spin Hall effect is not accompanied by transverse voltage in homogeneous systems on average.
Applications of the method of hybrid spin-noise spectroscopy will help to understand  in various spin-orbit coupled systems.

This work has been partially supported by the University of South Carolina ASPIRE grant 13070-12-29502. The work at LANL was funded by DOE under Contract No. DE-AC52-06NA25396.

\bibliography{spin}

\end{document}